\documentclass{aastex}
\usepackage{emulateapj5}     
\makeatletter

\renewcommand{\cite}[1]{\citeyear{#1}}
\bibpunct{\hspace{-4pt}}{}{}{a}{}{}
\submitted{{\rm ApJL in press}}  

\begin{document}

\title{Tidal scattering of stars on supermassive black holes in galactic
centers}

\author{Tal Alexander\and Mario Livio}

\affil{Space Telescope Science Institute, 3700 San Martin Drive, Baltimore,
MD 21218; tal@stsci.edu, mlivio@stsci.edu}

\begin{abstract}
Some of the mass that feeds the growth of a massive black hole (BH)
in a galactic center is supplied by tidal disruption of stars that
approach it on unbound, low angular momentum orbits. For each star
that is disrupted, others narrowly escape after being subjected to
extreme tidal distortion, spin-up, mixing and mass-loss, which may
affect their evolution and appearance. We show that it is likely that
a significant fraction of the stars around massive BHs in galactic
centers have undergone such extreme tidal interactions and survived
subsequent total disruption, either by being deflected off their orbit
or by missing the BH due to its Brownian motion. We discuss possible
long-term observable consequences of this process, which may be relevant
for understanding the nature of stars in galactic centers, and may
provide a signature of the existence of massive BHs there. 
\end{abstract}

\keywords{black hole physics --- galaxies: nuclei --- Galaxy: center --- Galaxy:
kinematics and dynamics --- stars: kinematics --- stars: rotation}

\section{Introduction}

\label{s:intro}

A massive black hole in a galactic nucleus grows by accreting matter
from its surroundings, some of it in the form of stars that approach
it on low angular momentum orbits (loss-cone orbits). When the BH
mass \( m \) is sufficiently small so that its tidal radius \( r_{t}\propto m^{1/3} \)
is larger than its Schwarzschild radius \( r_{S}\propto m \) (\( m\lesssim 10^{8}\, M_{\odot } \)
for the disruption of solar type stars), the star is tidally disrupted
before crossing the event horizon. The accretion of stellar debris
from such events may give rise to observable {}``tidal flares''
(Frank \& Rees \cite{Fra76}). Various authors estimated the rates,
timescales, luminosities and spectra of the flares (Ulmer, Paczy\'{n}ski
\& Goodman \cite{Ulm98}; Ulmer \cite{Ulm99}; Syer \& Ulmer \cite{Sye99};
Magorrian \& Tremaine \cite{Mag99}; Ayal, Livio \& Piran \cite{Aya00}).
There is to date only marginal evidence for the detection of tidal
flares (e.g. Renzini et al. \cite{Ren95}; Komossa \& Bade \cite{Kom99a};
Komossa \& Greiner \cite{Kom99b}).

In this \emph{letter} we propose another tidal process whereby the
BH may reveal its presence. We consider the fate of stars that narrowly
escape tidal disruption by the central BH after being subjected to
extreme tidal distortion, spin-up, mixing and mass-loss during periastron
passage. The total mass in such stars is comparable to that in stars
that are disrupted, and therefore also to \( \left( f_{t}/f_{m}\right) m \),
where \( f_{t} \) is the mass fraction of the BH that originates
in tidally disrupted stars, and \( f_{m} \) is the fraction of a
disrupted star's mass that is ultimately accreted (\( f_{m}\lesssim 0.5 \),
e.g. Ayal et al. \cite{Aya00}). The ratio \( f_{t}/f_{m} \) is significant
for a low mass BH (\( m\lesssim 10^{7}\, M_{\odot } \)) that accretes
from a low-density galactic nuclear core, where mass loss from stellar
collisions is small. Murphy, Cohn \& Durisen (\cite{Mur91}) found
\( f_{t}/f_{m}\sim \! 0.15 \) in Fokker-Planck simulations of the
growth of a low mass BH in a galactic nucleus (their model 4B), assuming
polytropic stellar models for estimating the collisional mass loss.
More recently, Freitag \& Benz (2001, in preparation) found \( f_{t}/f_{m}\sim \! 0.25 \)
for the same galactic nucleus model using Monte-Carlo simulations
with more realistic stellar models, which are more centrally concentrated
than polytropes, and therefore lose less mass in collisions. They
also find \( f_{t}/f_{m} \) as high as \( \sim \! 0.65 \) for low-density
nuclei where the stellar mass function is weighted toward low-mass
stars.

Dynamical analyses of the scattering of stars into loss-cone orbits
(Lightman \& Shapiro \cite{Lig77}; see Magorrian \& Tremaine \cite{Mag99}
for numeric examples) show that tidally disrupted stars in galactic
nuclei are typically on slightly unbound orbits relative to the BH
and that they are predominantly scattered into the loss-cone from
orbits at the radius of influence of the BH, \( r_{h}\equiv \left. Gm\right/ \sigma ^{2} \),
where \( \sigma  \) is the velocity dispersion far from the BH. The
stellar mass that is enclosed within \( r_{h} \) is comparable to
\( m \). The scattering timescale is shorter than the dynamical timescale,
and so the stars are scattered in and out of the loss-cone several
times during one orbital period. Following the first close passage
near the BH the stars are on very eccentric orbits with apoastron
\( \lesssim 2r_{h} \), and so there is a considerable chance that
they will be scattered again out of the loss-cone before their next
close passage, and thus avoid eventual orbital decay and disruption.

\begin{figure*}
{\centering \resizebox*{!}{0.3\textheight}{\includegraphics{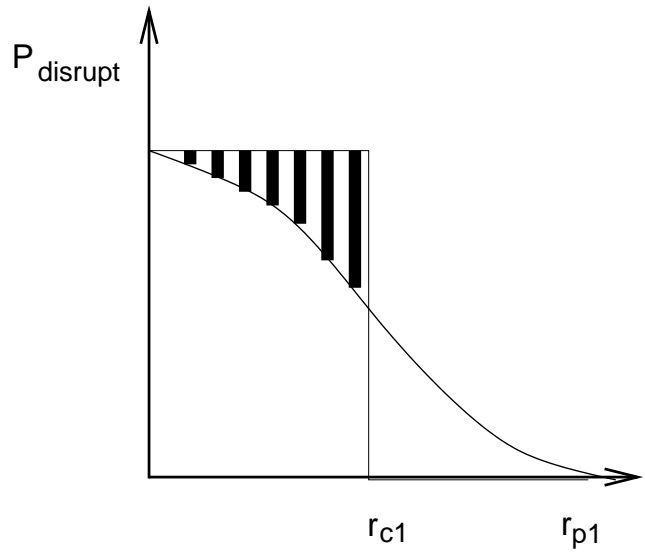}} \par}

\caption{\label{fig:Ps}A schematic representation of the change in the disruption
probability of a tidally disturbed star as function of the first peri-distance
\protect\protect\( r_{p1}\protect \protect \) when stochastic processes
are included. For an isolated 2 body system (i.e. no other stars present)
there exists a critical first peri-distance \protect\protect\( r_{c1}\protect \protect \)
such that a star passing at \protect\protect\( r_{p}<r_{c1}\protect \protect \)
will ultimately be tidally disrupted, and conversely, a star passing
at \protect\protect\( r_{p}>r_{c1}\protect \protect \) will survive
subsequent close passages. When processes such as scattering and the
BH Brownian motion are included, some of the stars with \protect\protect\( r_{p}<r_{c1}\protect \protect \)
will survive (hatched area).}
\end{figure*}

The survival probability is further increased by the Brownian motion
of the BH relative to the dynamical center of the stellar system.
In a system composed of stars with a typical mass \( M \) and radius
\( R \) and a low mass BH, which evolve in an initially constant
density core of radius \( r_{c}\sim \! r_{h} \), the amplitude of
the Brownian fluctuations is much larger than the tidal radius, \( \left\langle \Delta r\right\rangle /r_{t}\sim \! \left( r_{c}/R\right) \left( M/m\right) ^{5/6}\gg 1 \)
(Bahcall \& Wolf \cite{Bah76}). The Brownian motion proceeds on the
dynamical timescale of the core, which is comparable to the orbital
period of the tidally disturbed stars. While these results strictly
apply only if the BH is embedded in an isothermal system, which is
not the case in the GC, they are expected to hold generally to within
an order of magnitude. The orbits of the tidally scattered stars take
them outside of \( r_{h} \), where they are no longer affected by
the relative shift between the BH and the stellar mass. Therefore,
on re-entry into the volume of influence, their orbit will not bring
them to the same peri-distance from the BH. Both the random motion
of the BH and the scattering off the loss-cone by two-body interactions
are expected to increase the survival fraction \( f_{s} \) to a significantly
high value (Fig. \ref{fig:Ps}). More detailed calculations, which
integrate over the orbital distribution, are required to confirm these
qualitative arguments.

The cross section for a close passage with a peri-distance \( \leq r_{p} \)
scales as \( r_{p} \) for stars on nearly parabolic orbits (Hills
\cite{Hil75}; Frank \cite{Fra78}). It then follows that the number
of stars on nearly loss-cone orbits (\( r_{t}\lesssim r_{p}\lesssim 2r_{t} \))
is comparable to the number of disrupted stars, and so the mass fraction
associated with stars in the volume of influence that have undergone
extreme tidal disturbance, \( \left( f_{t}/f_{m}\right) f_{s} \),
is significant.

\section{Model}

\label{s:model}

The tidal interaction between the star and the BH is described in
the reduced mass system where the star, of mass \( M \) and radius
\( R \), is stationary and the BH approaches it on an unbound orbit
with a peridistance \( r_{p} \). The tilde symbol is used to denote
quantities expressed in the units where \( G=M=R=1 \). In these units
\( \widetilde{\Omega }=1 \) is the centrifugal break-up angular frequency
and \( \widetilde{E}=1 \) is the star's binding energy up to a factor
of order unity. High rotation is the longest lasting dynamical effect
of a hyperbolic tidal interaction that does not end in capture, and
so we will use the spin-up of the star by the tidal encounter, \( \Delta \widetilde{\Omega } \),
as a measure of the long-term effects on the star. A general treatment
of spin-up by hyperbolic encounters can be found in Alexander \& Kumar
(\cite{Ale01}), where it is applied to star-star tidal interactions.
Here we apply the formalism to the star-BH interaction.

The tidal disruption radius for slightly hyperbolic orbits is \( \widetilde{r}_{t}\simeq \widetilde{m}^{1/3} \)
in the limit \( \widetilde{m}\gg 1 \). It is convenient to parameterize
the tidal interaction in terms of the the penetration factor \( \beta =\widetilde{r}_{t}/\widetilde{r}_{p} \).
The linear tidal coupling coefficients \( T_{l} \) are functions
of the parameter \( \eta =\widetilde{r}^{3/2}_{p}\left/ \sqrt{1+\widetilde{m}}\right.  \),
where \( \eta \simeq \beta ^{-3/2} \) in the limit \( \widetilde{m}\gg 1 \),
and of the eccentricity of the orbit \( e=1+2\widetilde{r}_{p}\widetilde{E}_{o}/\widetilde{m} \),
where \( \widetilde{E}_{o} \) is the orbital energy of the 2-body
reduced mass system. The spin-up due to the transfer of angular momentum
from the orbit by the tidal interaction is related to the energy transfer
by \( \Delta E=I\Omega _{p}\Delta \Omega  \), where \( \Omega _{p} \)
is the relative angular velocity at periastron, \( I \) is the star's
moment of inertia, and rigid body rotation is assumed (Goldreich \&
Nicholson \cite{Gol89}; Kumar \& Quataert \cite{Kum98}). On parabolic
orbits \( \Delta  \)\( \Omega  \) is independent of \( m \) to
leading order in the multipole expansion, and can be expressed as
(e.g. Press \& Teukolsky \cite{Pre77}) \begin{equation}
\Delta \widetilde{\Omega }\simeq \frac{T_{2}(\beta ^{-3/2})}{\sqrt{2}\widetilde{I}}\beta ^{9/2}\, .
\end{equation}
 The linear analysis of the spin-up breaks down as \( \widetilde{r}_{p} \)
decreases and \( \Delta \widetilde{\Omega } \) approaches 1. At first
\( \Delta \widetilde{\Omega } \) increases faster than predicted
by the linear analysis, but then it peaks at the onset of mass-loss
as the ejecta carry away the angular momentum. In the results presented
here we do not correct for the non-linear effects, and so formally
very high values of \( \Delta \widetilde{\Omega } \) should be understood
as indicating any combination of mass-loss, mixing or spin-up effects.
The tidal coupling coefficients can be calculated numerically for
any given stellar model. In our analysis we will use two stellar models:
a \( n=3/2 \) ideal gas polytrope, and a solar model (Christensen-Dalsgaard
et al. \cite{Chr96}).

We demonstrate the tidal scattering effect by considering the fate
of a star on a circular orbit at an initial radius \( r_{i}=r_{h} \)
that loses angular momentum, but not energy, in a 2-body interaction
and is deflected into a near loss-cone orbit with peri-distance \( r_{t}<r_{p}\ll r_{i} \).
We assume for simplicity that the stellar mass within \( r_{i} \)
equals the BH mass, that it is negligible within \( r_{p} \), and
that its distribution is proportional to \( r^{-\alpha } \) (\( \alpha \neq 2,\, 3 \)).
It then follows from energy conservation on orbits in a spherical
mass distribution that the orbital energy of the reduced mass system
at \( \widetilde{r}_{p} \) (i.e. excluding the potential of the stellar
cluster) is \begin{equation}
\widetilde{E}_{o}=\frac{1}{2-\alpha }\left( \frac{\widetilde{m}}{\widetilde{r}_{i}}\right) \, ,
\end{equation}
 and the eccentricity is \begin{equation}
e=1+\frac{2}{2-\alpha }\left( \frac{\widetilde{r}_{p}}{\widetilde{r}_{i}}\right) \, .
\end{equation}

\section{Results}

\label{s:results}

\begin{figure*}
{\centering \resizebox*{!}{0.3\textheight}{\includegraphics{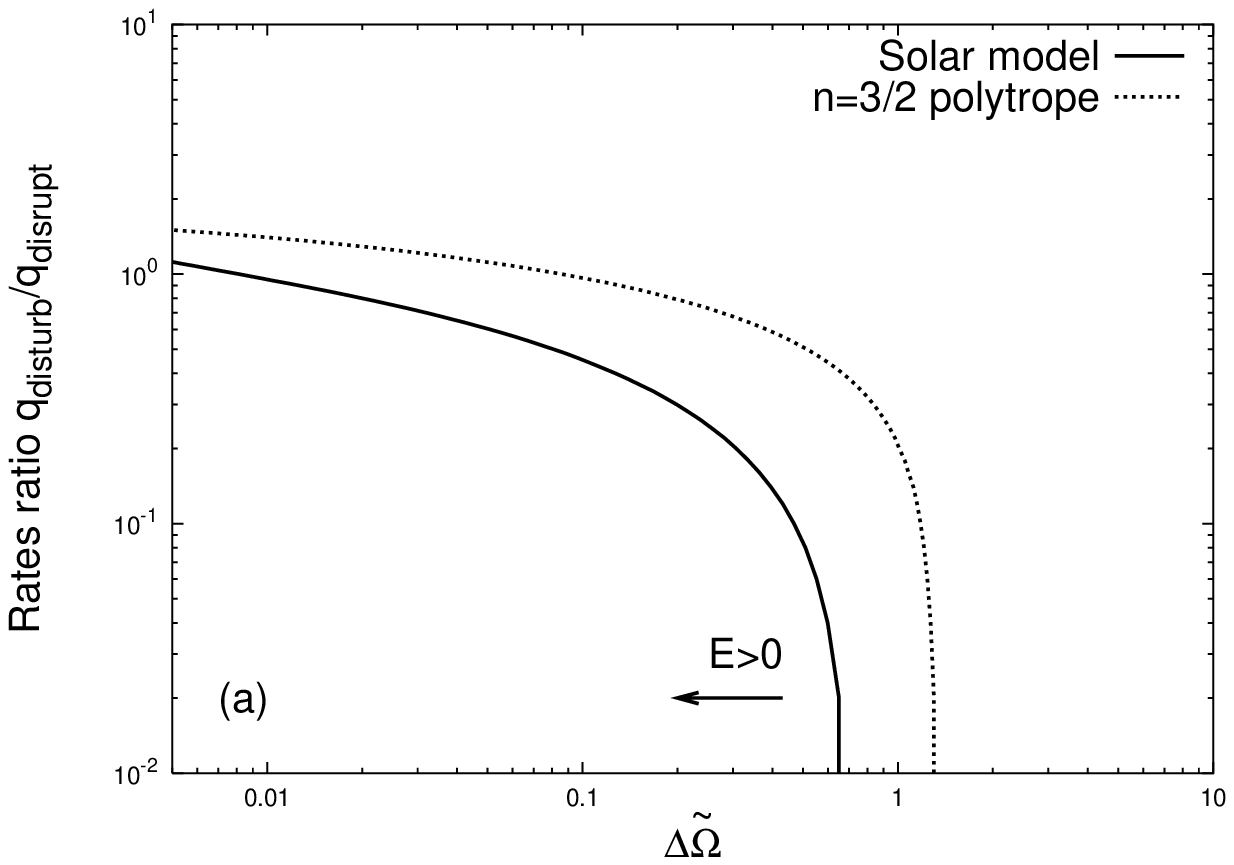}} \par}

{\centering \resizebox*{!}{0.3\textheight}{\includegraphics{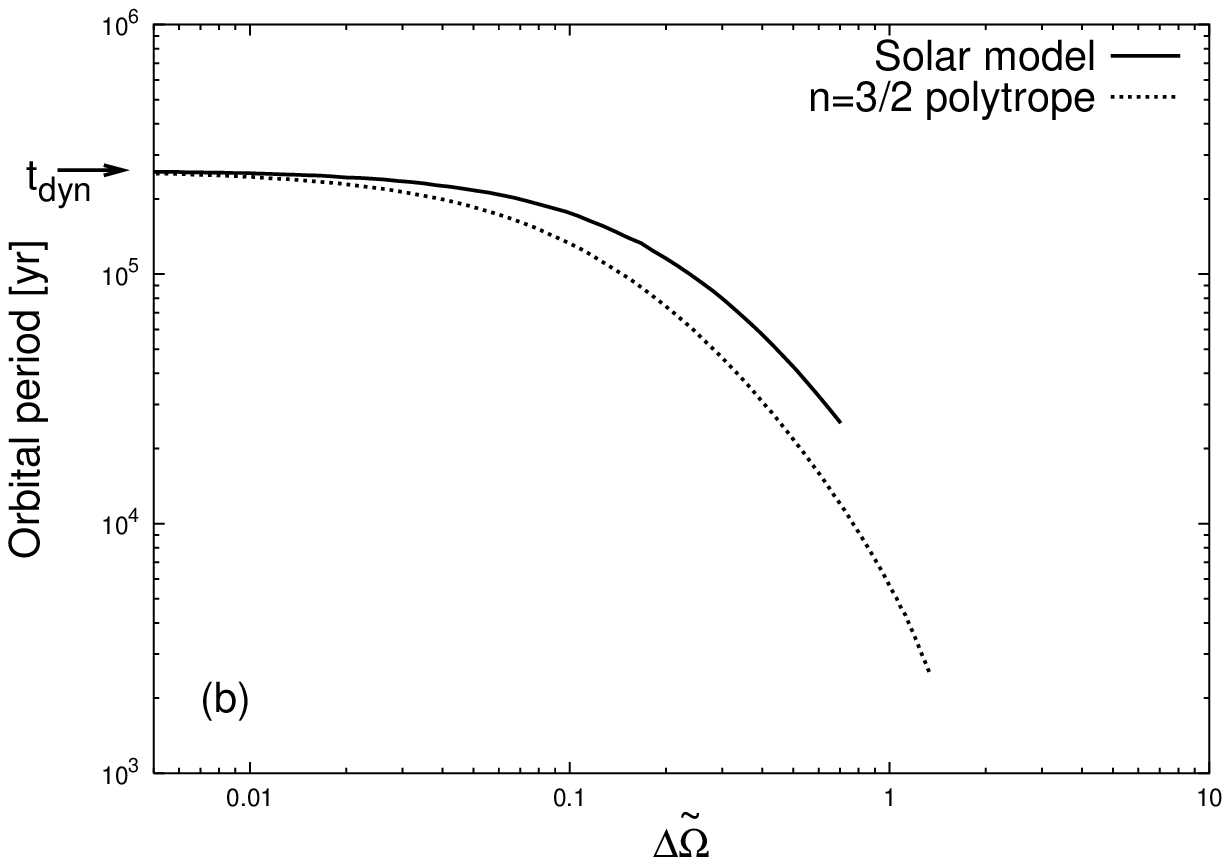}} \par}

{\centering \resizebox*{!}{0.3\textheight}{\includegraphics{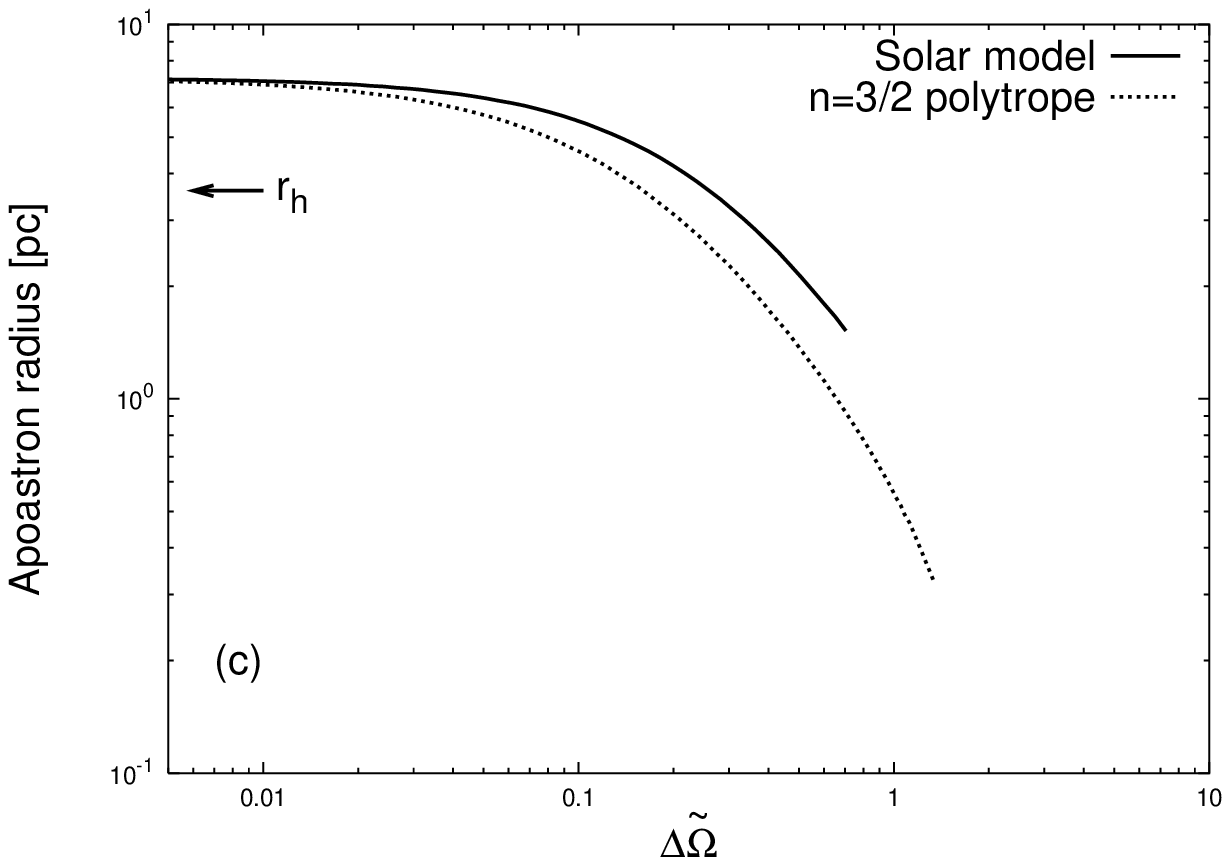}} \par}

\caption{\label{fig:dW}(a) The ratio between the rate of stars that undergo
extreme tidal disturbance and those that are tidally disrupted in
a model of the GC (\protect\protect\( M_{\mathrm{BH}}=3\times 10^{6}\, M_{\odot }\protect \protect \),
\protect\protect\( r_{1}=r_{h}=3.6\, \mathrm{pc}\protect \protect \),
\protect\protect\( \alpha =1.5\protect \protect \)). The arrow marks
the range of \protect\protect\( \Delta \widetilde{\Omega }\protect \protect \)
where the star remains unbound to the BH after the first periastron
passage. (b) The orbital period after the first periastron passage.
The arrow marks the orbital period at the radius of influence. (c)
The apoastron distance after the first periastron passage. The arrow
marks the BH radius of influence.}
\end{figure*}

We apply our model to the \( 3\times 10^{6}\, M_{\odot } \) BH in
the Galactic Center (GC), which is of particular interest because
present-day observations can already resolve individual stars very
close to the BH (Genzel et al. \cite{Gen97}; Ghez et al. \cite{Ghe98}).
The BH in the GC is surrounded by a 2-body relaxed cusp with \( \alpha \sim \! 1.5 \)
and has a radius of influence of \( r_{h}\sim \! 3.6\, \mathrm{pc} \)
(Alexander \cite{Ale99}). For the BH in the GC, \( r_{t}>10r_{S} \)
for a solar type star, and so general relativistic corrections can
be safely neglected (Laguna et al. \cite{Lag93}). Figure \ref{fig:dW}
shows the cumulative values of the ratio between the rate of near
loss-cone encounters (\( r_{p}>r_{t}) \) and the total rate of disruptive
encounters, the orbital period after the first periastron passage,
and the apoastron distance after the first periastron passage, all
as functions of \( \Delta \widetilde{\Omega } \). The orbits have
\( \widetilde{E}_{o}=0.04 \), but are still effectively parabolic,
with \( 1<e<1+10^{-5} \). For the solar stellar model, the orbital
energy remains positive after the first periastron passage for all
values of \( \Delta \widetilde{\Omega }<\Delta \widetilde{\Omega }_{b}=0.43 \)
(\( r_{p}/r_{t}>1.13 \)), which corresponds to unbound \emph{and}
tidally disturbed stars (defined here as having \( 1/2<\Delta \widetilde{\Omega }/\Delta \widetilde{\Omega }_{b}<1 \))
of total mass \( 0.16\left( f_{t}/f_{m}\right) f_{s}m \). For the
polytropic stellar model, which is less centrally concentrated and
therefore more easily perturbed by the tidal interaction, the orbital
energy remains positive after the first periastron passage for all
values of \( \Delta \widetilde{\Omega }<0.28 \) (\( r_{p}/r_{t}>1.70 \)),
which corresponds to unbound tidally disturbed stars of total mass
\( 0.19\left( f_{t}/f_{m}\right) f_{s}m \). The plots of the orbital
period and apoastron radius confirm that the qualitative arguments
presented in \S\ref{s:intro} hold for all but the closest periastron
passages.

\section{Discussion}

\label{s:discuss}

We have argued that a significant fraction of the stars around a supermassive
black hole in a galactic center experience considerable spin-up, due
to extreme tidal interactions with the central BH. Unlike the star--star
tidal interactions considered by Alexander \& Kumar (\cite{Ale01}),
which also lead to high rotation, this mechanism does not require
an extremely high stellar density. We further note that the tidally
disturbed stars will remain in the galactic center even after tidal
scattering ceases in the later phase of the BH growth when its mass
exceeds \( \sim \! 10^{8}\, M_{\odot } \) and \( r_{S}>r_{t} \).
Assuming \( f_{s}\sim \! 0.5 \), \( f_{t}/f_{m}\sim \! 0.25 \) and
solar type stars, and considering only orbits that are unbound to
the BH after the first periastron passage, we estimate that the Galactic
Center contains \( 10^{4-5} \) stars that have survived an extreme
tidal interaction with the central BH.

We now examine briefly potential observational signatures of such
interactions. The effects of rotation on stellar evolution have been
studied extensively by a number of authors (see e.g. Maeder \& Meynet
\cite{Mae00}; Heger \& Langer \cite{Heg00}, and references therein).
Here we want to concentrate in particular on a few observable effects:
(1) The luminosity, (2) abundance anomalies, and (3) colors. Rotationally
induced mixing results in higher bolometric luminosities for a given
mass, due to the increase in the mean molecular weight \( \mu  \),
and in the convective core size. Mixing also brings the products of
hydrogen burning to the stellar surface. Thus, one can expect enrichment
in \( ^{4} \)He, \( ^{14} \)N, \( ^{13} \)C, and \( ^{26} \)Al,
and depletion in \( ^{12} \)C, \( ^{16} \)O, \( ^{15} \)N. Combining
these two effects, one of the consequences of high rotation rates
is stronger abundance anomalies in more luminous stars. Interestingly,
these were precisely the effects observed by Carr, Sellgren \& Balachandran
(\cite{Car00}), in their abundance measurements of the M supergiant
IRS 7 in the Galactic center. The low C and O and high N abundances
indicate that IRS 7 has undergone mixing in excess of the values predicted
by standard models (or observed in other supergiants like \( \alpha  \)Ori).
In fact, the observed abundances have prompted Carr et al. to conclude
that {}``extra mixing induced by rapid rotation may indeed be the
fundamental difference between the evolution of massive stars in the
Galactic center and those elsewhere in the Galaxy.'' The third effect
is relevant particularly to stars that would have otherwise (with
no tidal spin-up) been red supergiants or Asymptotic Giant Branch
(AGB) stars. The envelopes of such stars (particularly the latter)
are relatively loosely bound to their cores. Consequently, even modest
amounts of tidal interaction and spin-up lead to a loss of all or
a part of the envelope (e.g. Livio \cite{Liv94}). The stripped, more
compact star will thus become blue. In fact, this mechanism has been
proposed to explain the existence of a strange group of very blue
stars in the globular cluster M15 (De Marchi \& Paresce \cite{DeM96}),
and the colors of Sk~\( -69^{\circ }\, 202 \), the progenitor of
SN1987A (e.g. Woosley \cite{Woo88}). Our calculations therefore predict
that the stellar population in galactic centers will exhibit a paucity
in red giants, and will rather be relatively rich in blue stars. Concentrations
of blue stars around the central BH are found in both the Galactic
Center (Genzel et al. \cite{Gen97}) and in M31 (Lauer et al. \cite{Lau98}).
Many of the blue stars are predicted to be on elongated ({}``radial'')
orbits. Observations of the Sgr A\( ^{\star } \) stellar cluster
in the Galactic center appear to be consistent with this prediction
(Genzel et al. \cite{Gen00}), although the orbital solutions are
still quite uncertain (Ghez et al. \cite{Ghe00}).

\acknowledgements{We thank Marc Freitag, Andrea Ghez, Tod Lauer and John Magorrian
for useful discussions. TA is grateful for the hospitality of the
Aspen Center for Physics, where this work was initiated.}

\end{document}